\documentclass{ws-procs9x6}

\usepackage{amssymb}
\usepackage{graphicx}

\newcommand\Journal[4]{\textit{#1}~\textbf{#2}, #3~(#4)}
\newcommand\PRD{Phys.~Rev.~D}
\newcommand\PLB{Phys.~Lett.~B}

\graphicspath{
  {/afs/desy.de/user/b/beckmann/talks/qcd.virginia.apr2002/buidln/}
  {/afs/desy.de/user/b/beckmann/talks/dis2001/buidln/}
  }

\begin{document}


\title{Recent Results on the Helicity Structure of the Nucleon from HERMES}

\author{Marc Beckmann}

\address{on behalf of the HERMES Collaboration\\[2ex]
  Deutsches Elektronen--Synchrotron DESY,
  D--22603 Hamburg, Germany\\
  E-mail: marc.beckmann@desy.de}

\maketitle


\protect\abstracts{
  The HERMES experiment has measured double spin asymmetries of
  inclusive and semi--inclusive cross sections for the production of
  charged hadrons in deep--inelastic scattering of 
  polarised positrons on 
  polarised hydrogen and deuterium targets, in the kinematic range
  0.023 $< x <$ 0.6, and 1 GeV$^2 < Q^2 <$ 15 GeV$^2$.  For the data
  taken on the deuterium target, a RICH detector provides the complete
  identification of charged pions and kaons.  From the inclusive
  measurements on deuterium, the polarised structure function $g_1^d$
  has been extracted with high precision.  Together with the
  semi--inclusive measurements, quark polarisations are extracted
  separately for the $u$, $\bar{u}$, $d$, $\bar{d}$, and $(s +
  \bar{s})$ flavours in a LO QCD analysis. 
  Furthermore, a possible breaking of flavour symmetry in the
  polarised light sea, $\Delta\bar{u} - \Delta\bar{d}$, was measured for
  the first time and found to be consistent with zero within the
  experimental precision.
  }


\section{Formalism of the Analysis}

Cross section asymmetries measured in doubly polarised deep--inelastic
scattering (DIS) provide a useful tool to probe the helicity structure
of the nucleon.  Under the assumption of a vanishing spin structure
function $g_2 = 0$, the experimentally measured asymmetry
$A_\parallel$ is related to the asymmetry $A_1$ in the virtual photon
-- nucleon reference frame by
\begin{equation}
  \label{eq:Apar_A1}
  \frac{A_\parallel}{D(1+\eta\gamma)} \: \simeq \: A_1 \: \simeq \:
  \frac{g_1}{F_1} \: .
\end{equation}
Here, $D$ is the depolarisation factor of the virtual photon, $g_1$
and $F_1$ are the polarised and the unpolarised nucleon structure
functions, respectively, and $\eta$ and $\gamma = \sqrt{Q^2/\nu^2}$
are kinematic factors.  Here, $\nu$ is the energy of the virtual
photon, $-Q^2$ is its four--momentum squared, and $x = Q^2/2M\nu$,
where $M$ is the nucleon rest mass.  Contributions from $g_2$, found
to be very small in measurements at SLAC [\refcite{lit:E155X:g2}], are
omitted.  A residual effect of the non--zero value of $g_2$ on the
extracted quark polarisations has been included in the systematic
uncertainty.

In leading order (LO) QCD the relation $g_1(x,Q^2) \: = \: \sum_q
e_q^2 \Delta q(x,Q^2)$ applies, where $\Delta q(x,Q^2) \equiv
q^+(x,Q^2) - q^-(x,Q^2)$ are the helicity quark distributions in the
nucleon.   Information on the charge squared weighted sum
over the quark helicity distributions can hence be obtained based on
measurements of $g_1$ according to Eq.~(\ref{eq:Apar_A1}).  Here,
$q^{+(-)}(x,Q^2)$ is the number density of quarks with spin
aligned parallel (anti--parallel) to the nucleon's spin.

The observation of coincident hadrons from the current fragmentation
region in semi--inclusive deep--inelastic scattering (SIDIS) processes
provides access to the flavour of the struck quark.  Assuming
factorisation, the SIDIS cross section asymmetry $A_1^h$ can be
written in LO QCD as
\begin{equation}
  \label{eq:A1_dq_q}
  A_1^h(x, Q^2) \: \stackrel{g_2=0}{\simeq} \:
  \frac{1 + R(x,Q^2)}{1 + \gamma^2} \: \times \:
  \frac{\sum_q e_q^2 \: \Delta q(x,Q^2) \:
    \int_{z_{\mathrm{min}}}^{z_{\mathrm{max}}} D_q^h(Q^2\!,z)
    \:{\mathrm d}z}{\sum_q e_q^2 \: q(x,Q^2) \:
    \int_{z_{\mathrm{min}}}^{z_{\mathrm{max}}}
    D_q^h(Q^2\!,z) \:{\mathrm d}z} \: .
\end{equation}
Here, $q(x,Q^2)$ denote the unpolarised parton densities (PDFs) and
$R(x,Q^2) = \sigma_L / \sigma_T$ is the photo--absorption cross
section ratio for longitudinal and transverse virtual photons.  In
this analysis, the fragmentation functions $D_q^h(Q^2\!,z)$ are
integrated over the range $z_{\mathrm{min}} = 0.2$ to
$z_{\mathrm{max}} = 0.8$ in $z = E_h/\nu$, which is the fraction of
the virtual photon's energy carried by the hadron of type $h$.
The upper cut on $z \leq 0.8$ eliminates exclusive events from the
sample.

For a set of measured inclusive and semi--inclusive asymmetries on
different targets, after integrating over the range in $Q^2$ in each 
$x$--bin, Eq.~(\ref{eq:A1_dq_q}) can be rewritten in matrix form
\begin{equation}
  \label{eq:puri}
  \vec{\mathsf A}(x) \: = \: P(x) \: \vec{\mathsf Q}(x) \: .
\end{equation}
The vectors $\vec{\mathsf A}(x)$ and $\vec{\mathsf Q}(x)$ contain the
measured asymmetries and the polarisations $\Delta q/q$ of the
different quark flavours to be extracted, respectively.  The elements
of the matrix $P(x)$ depend on the fragmentation functions, the PDFs,
the cross section ratio $R(x,Q^2)$, and on the relative fluxes of
hadrons originating from the different nucleons in case of deuterium.
$P(x)$ also contains the influence from the limited acceptance of the
spectrometer.  The fragmentation was modelled in the LUND string model
implemented in the JETSET 7.4~[\refcite{lit:JETSET}] package.  In
order to obtain a reliable description of the fragmentation at HERMES
energies, the LUND string breaking parameters have been tuned to fit
the hadron multiplicities measured at HERMES.  As an estimate of the
associated systematic uncertainties, the analysis was also carried out
with an alternative fit to the measured multiplicities and with a
default LUND parameter setting for the PETRA $e^+e^-$ collider.  The
resulting variations of the extracted quark polarisations were added
in quadrature to the other sources of systematic uncertainties.  For
the unpolarised PDFs the parameterisations of
Ref.~[\refcite{lit:cteq5}] (CTEQ5LO) were used.

\begin{figure}[t]
  \begin{minipage}{\textwidth}
    \begin{minipage}[t]{0.48\textwidth}
      \includegraphics[width=\textwidth]{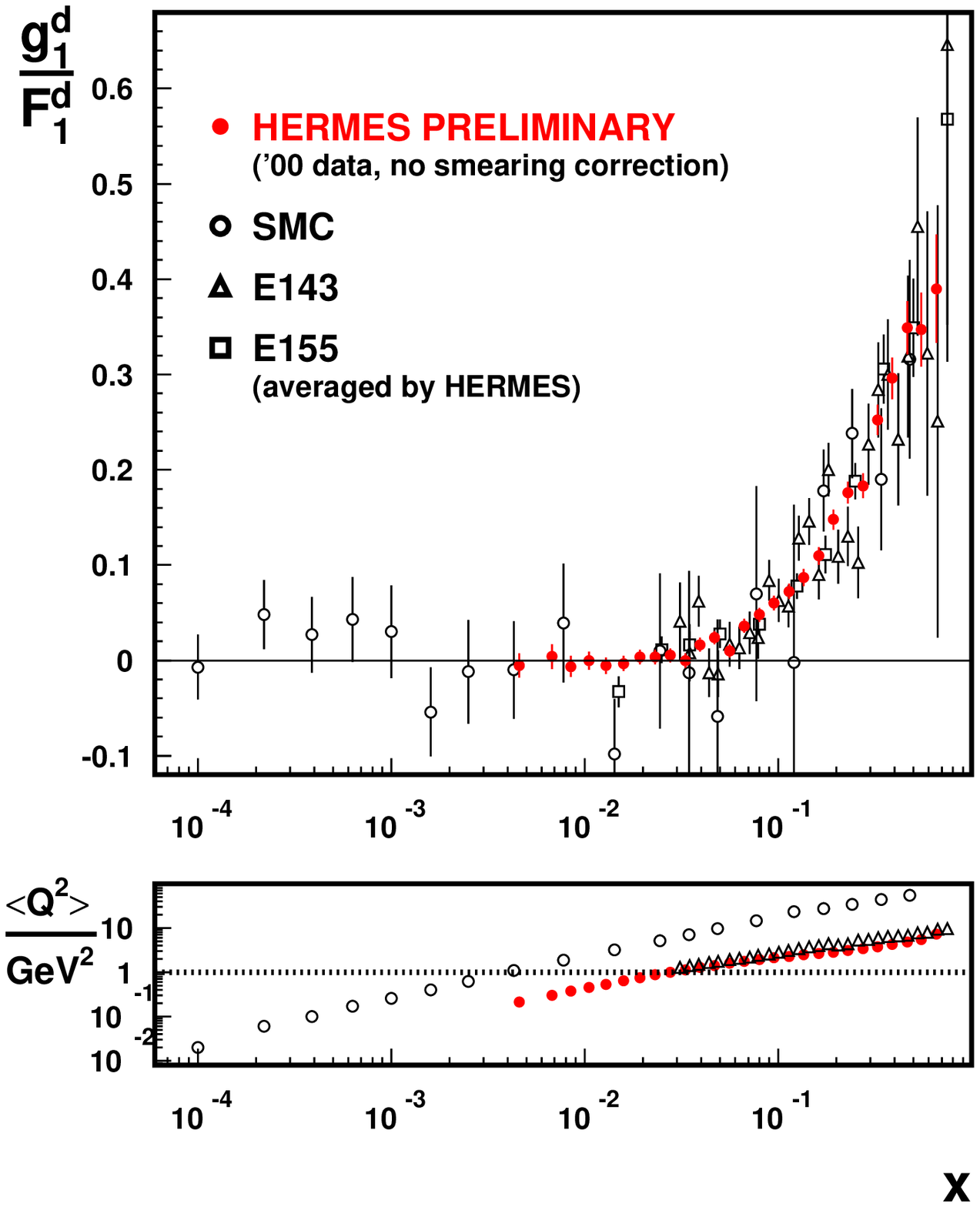}
    \end{minipage}
    \hspace*{0.02\textwidth}
    \begin{minipage}[t]{0.48\textwidth}
      \raisebox{0.6mm}{
      \includegraphics[width=\textwidth]{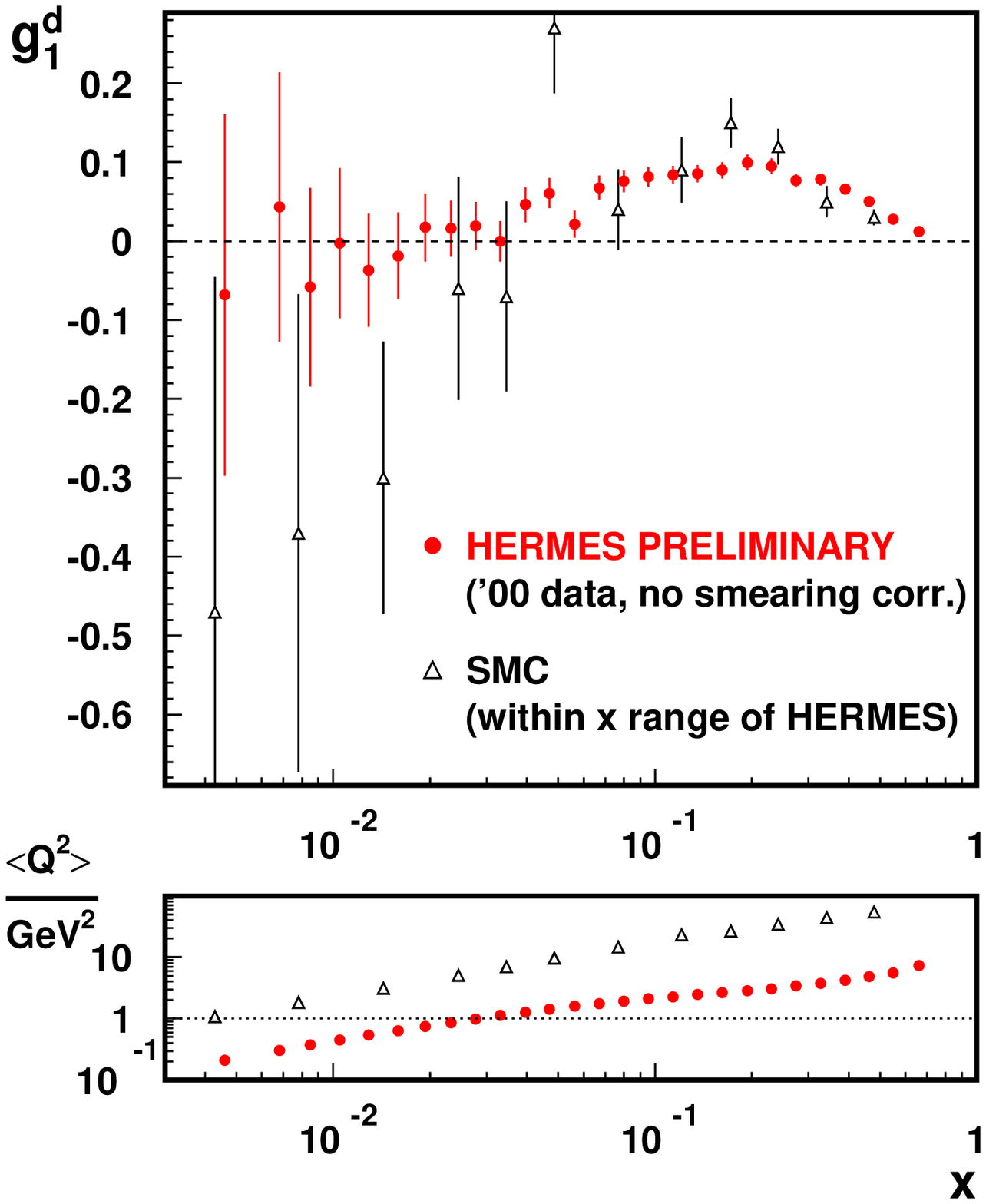}}
    \end{minipage}
  \end{minipage}
  \caption{\label{fig:g1d} Preliminary HERMES results for $g_1^d/F_1^d$
    and $g_1^d$ as a function of $x$.}
  \vspace*{-1.5ex}
\end{figure}

\begin{figure}[t]
  \vspace*{-5mm}
  \begin{minipage}{\textwidth}
    \begin{minipage}[t]{0.5\textwidth}
      \hspace*{-0.05\textwidth}
      \includegraphics[width=1.03\textwidth]{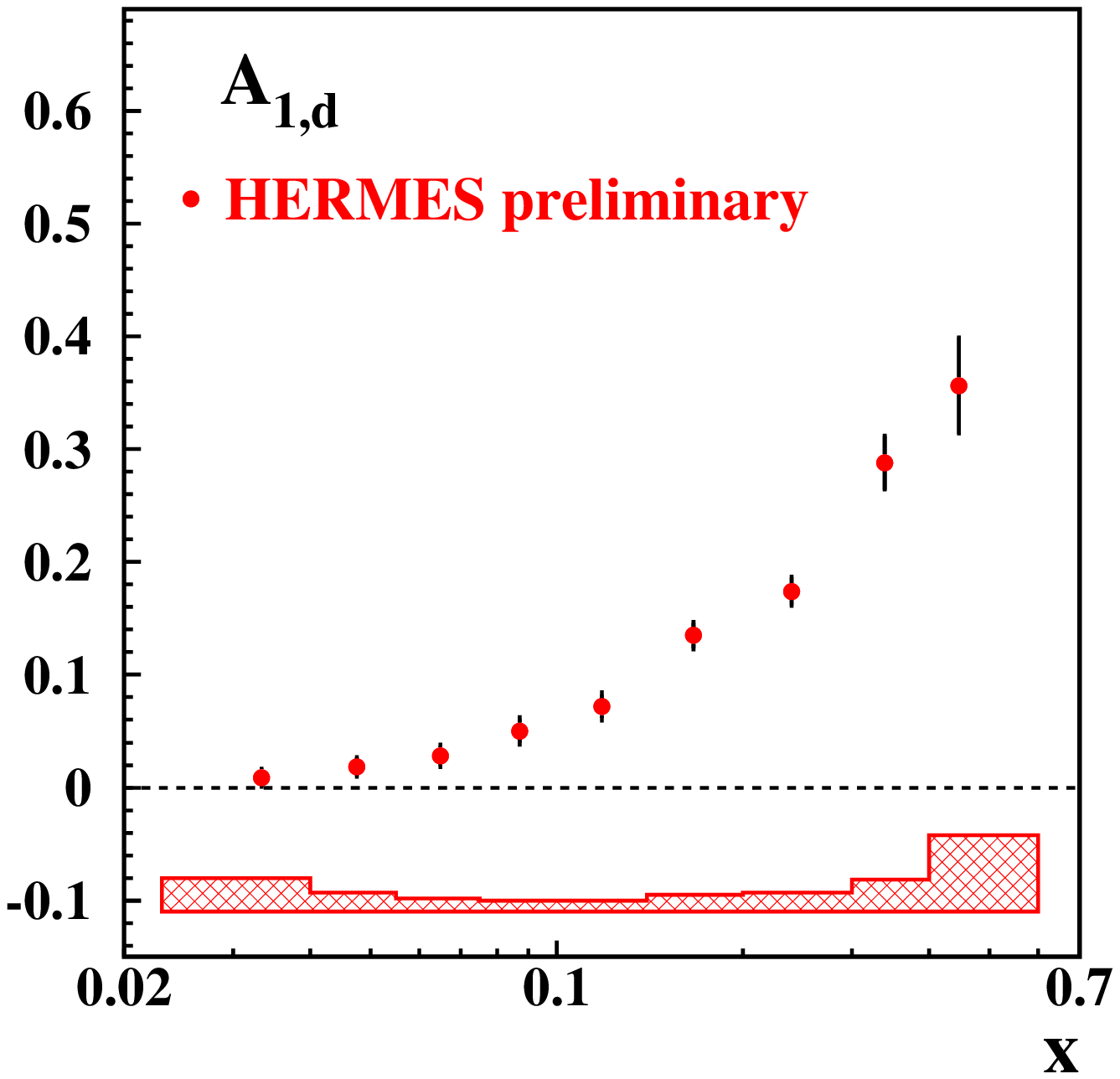}
    \end{minipage}
    \hspace*{-0.02\textwidth}
    \begin{minipage}[t]{0.5\textwidth}
      \hspace*{-0.03\textwidth}
      \includegraphics[width=1.03\textwidth]{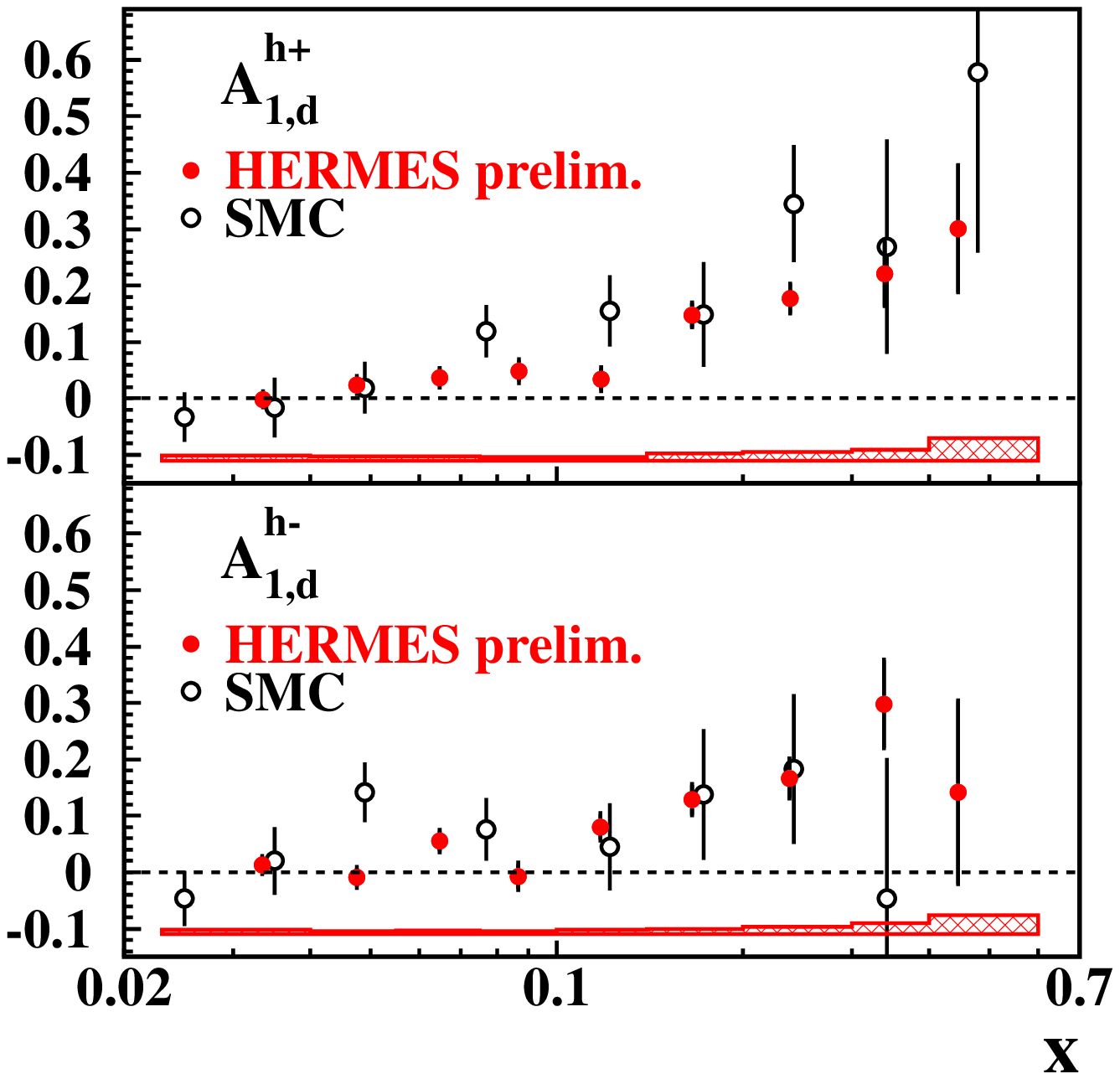}
    \end{minipage}
    \vspace*{-10mm}
  \end{minipage}
  \begin{minipage}{\textwidth}
    \begin{minipage}[t]{0.5\textwidth}
      \hspace*{-0.05\textwidth}
      \includegraphics[width=1.03\textwidth]{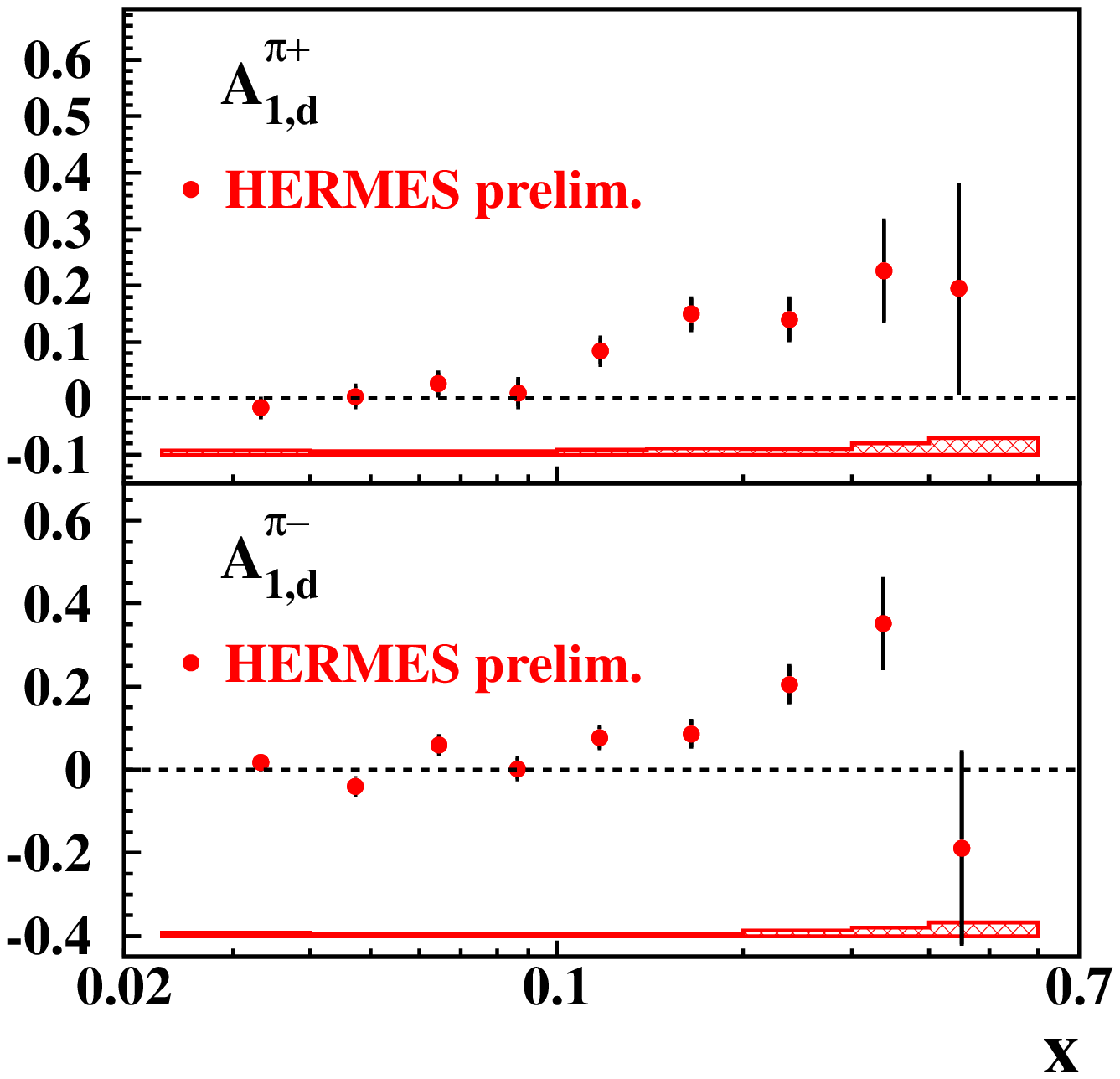}
    \end{minipage}
    \hspace*{-0.02\textwidth}
    \begin{minipage}[t]{0.5\textwidth}
      \hspace*{-0.03\textwidth}
      \includegraphics[width=1.03\textwidth]{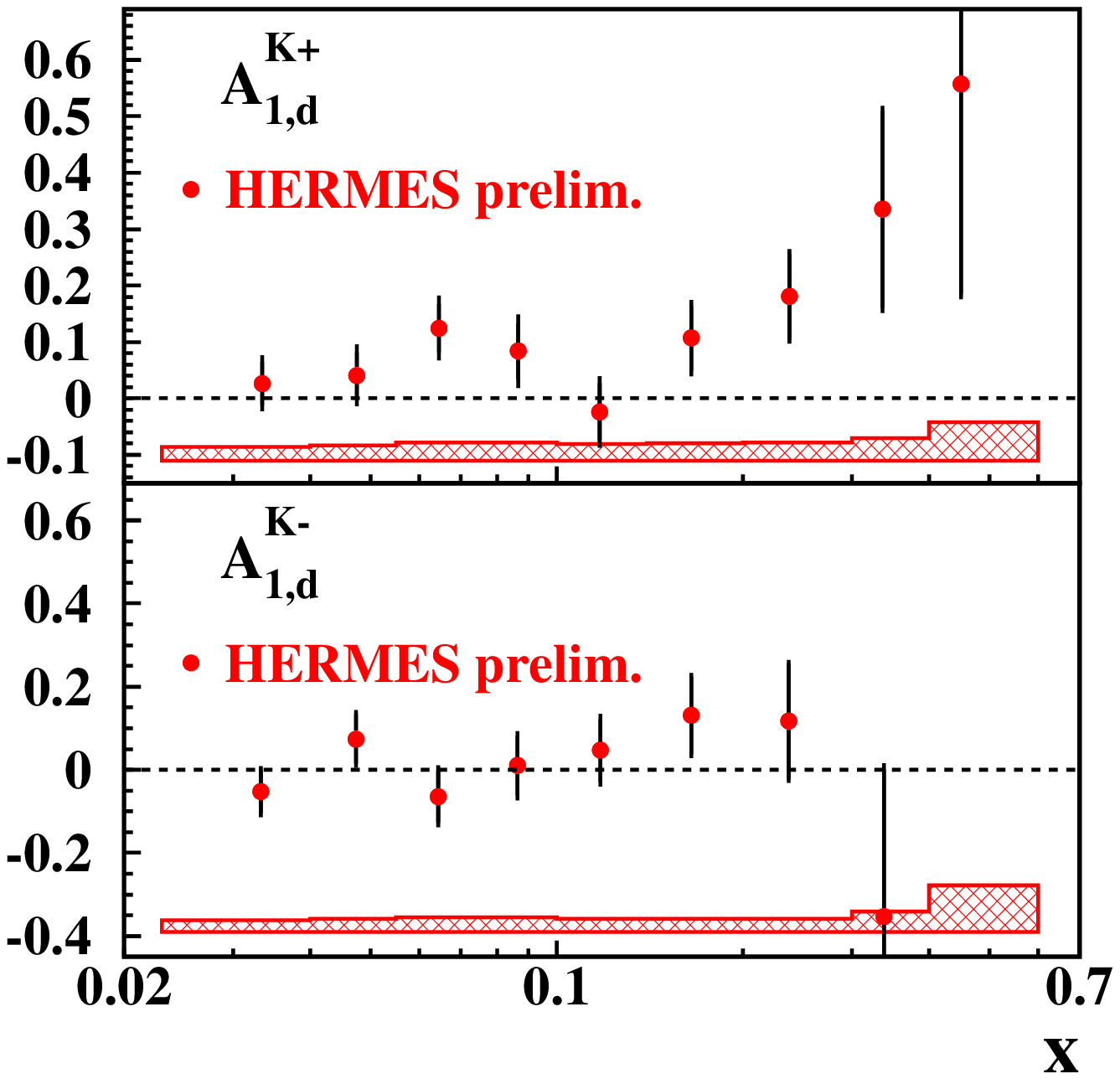}
    \end{minipage}
  \end{minipage}
  \vspace*{-4mm}
  \caption{\label{fig:A1d} Preliminary HERMES results on inclusive and
    semi--inclusive DIS cross section asymmetries $A_{1,d}(x)$ on a
    polarised deuterium target.
    }
  \vspace*{-1.5ex}
\end{figure}

The flavour decomposition of the quark polarisations was obtained by
solving Eq.~(\ref{eq:puri}) for the vector $\vec{\mathsf Q}(x) \: = \:
\left( \frac{\Delta u}{u}, \frac{\Delta\bar{u}}{\bar{u}}, \frac{\Delta
    d}{d}, \frac{\Delta\bar{d}}{\bar{d}}, \frac{\Delta s}{s} \equiv
  \frac{\Delta\bar{s}}{\bar{s}} \right)$.  In contrast to earlier
analyses [\refcite{lit:SMC:dq,lit:HERMES:dq}], the only remaining
symmetry assumption is $\Delta s(x)/s(x) = \Delta\bar{s}(x)/\bar{s}(x)$.


\section{Experimental Results}

The HERMES experiment~[\refcite{lit:HERMES:spectr}]~uses the longitudinally
polarised 27.6 GeV positron beam of the HERA storage ring at DESY,
incident on longitudinally polarised pure atomic gas targets of
$^3$He, hydrogen or deuterium.  A large acceptance forward
spectrometer with good particle identification detects the scattered
beam positrons together with coincident final state hadrons.
A RICH detector installed in 1998 provides the identification of pions,
kaons and protons over a wide momentum range for the data on the polarised
deuterium target taken since then.


\subsection{Inclusive Measurements}

Fig.~\ref{fig:g1d} shows the preliminary HERMES data on the structure
function ratio $g_1^d/F_1^d$ and on the polarised structure function
$g_1^d$.  These very precise results are based on 10 million
events taken on a polarised deuterium target in the kinematic range
$Q^2 > 0.1$ GeV$^2$ and $W^2 > 3.24$ GeV$^2$.  Despite the differences
in the average values of $\langle Q^2 \rangle$ per bin, one observes a
good consistency of the HERMES data on $g_1/F_1$ with an earlier
measurement of SMC at larger values of $\langle Q^2 \rangle$ and
measurements at SLAC, taken at similar average values of $Q^2$.  This
close agreement is a clear indication for very similar $Q^2$
evolutions for both the polarised and the unpolarised structure
functions over the range in $Q^2$ covered by experiments.


\subsection{Semi--Inclusive Measurements}

\begin{figure}[t]
  \vspace*{-5mm}
  \begin{minipage}{\textwidth}
    \begin{minipage}[t]{0.5\textwidth}
      \hspace*{-0.07\textwidth}
      \includegraphics[width=1.1\textwidth]{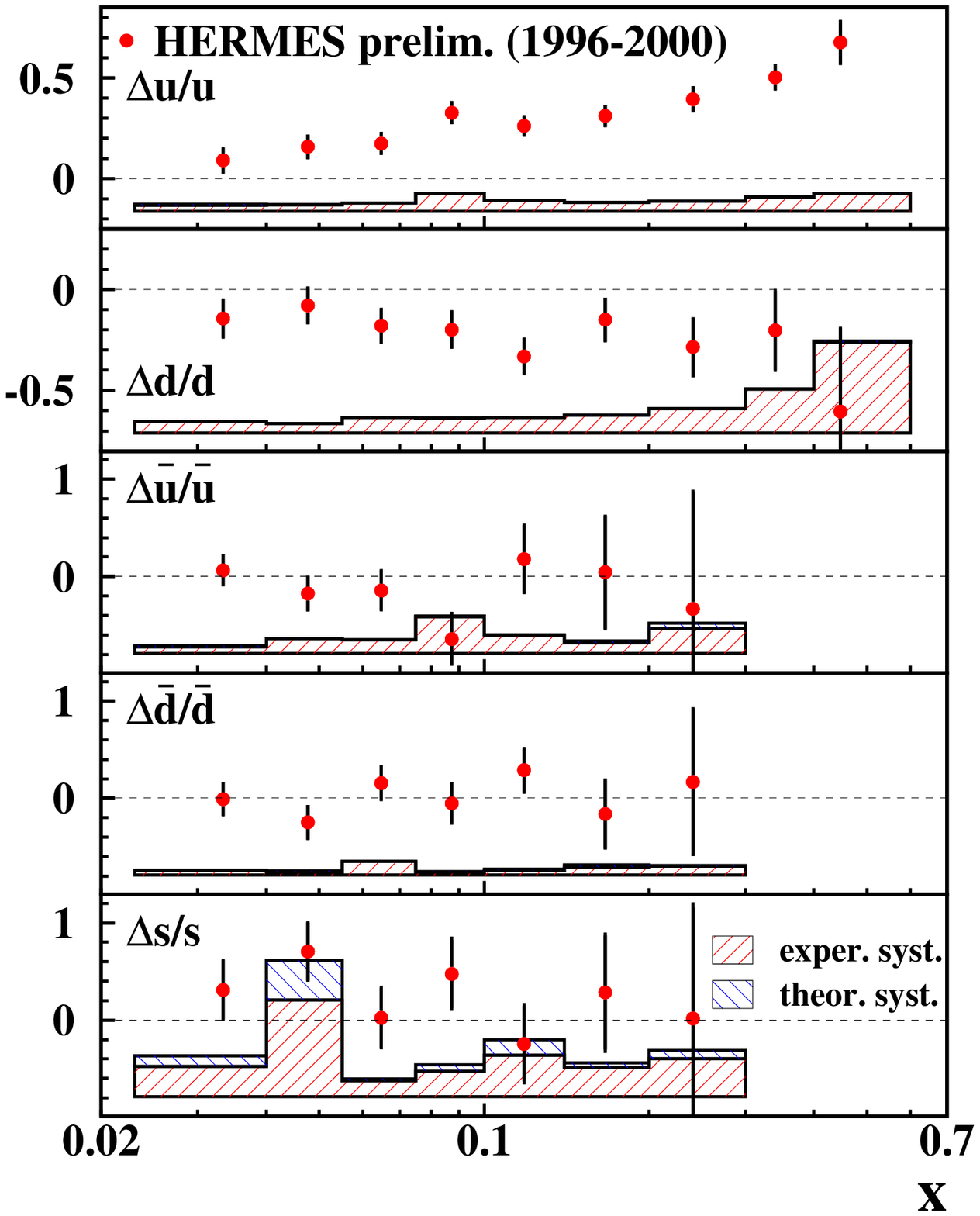}
    \end{minipage}
    \hspace*{-0.02\textwidth}
    \begin{minipage}[t]{0.5\textwidth}
      \hspace*{-0.05\textwidth}
      \includegraphics[width=1.1\textwidth]{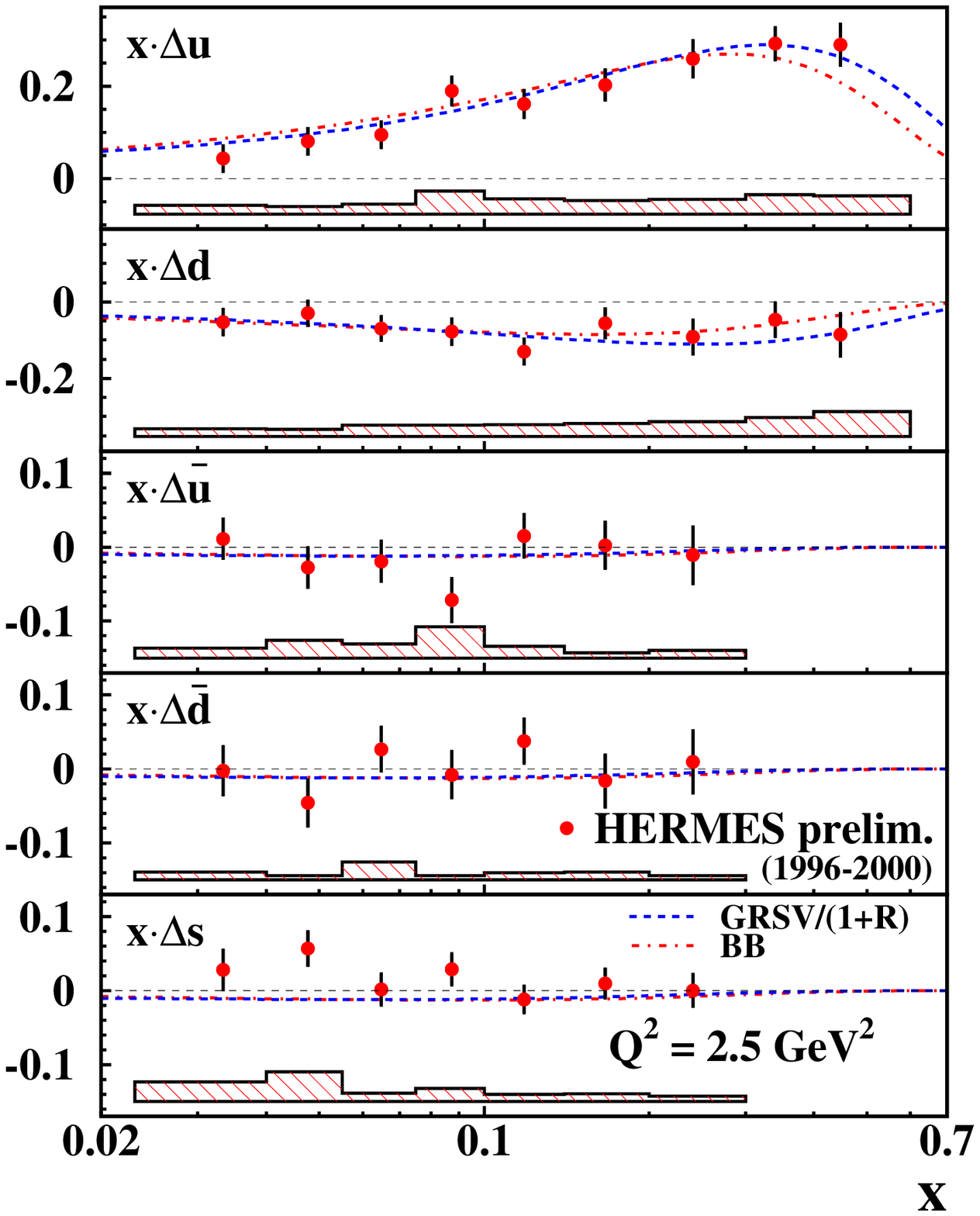}
    \end{minipage}
  \end{minipage}
  \vspace*{-4mm}
  \caption{\label{fig:dq5par} HERMES preliminary results on the polarisation
    of the $u$, $\bar{u}$, $d$, $\bar{d}$, and $s+\bar{s}$ quark
    flavours versus $x$ (left panel).  The right panel shows the
    results on the corresponding polarised quark distributions at a
    common scale of $Q^2 = 2.5$ GeV$^2$.  The dashed and the
    dotted--dashed lines are parameterisations from
    Refs.~[\protect\refcite{lit:grsv2000,lit:bbfit}].  The error bars
    are statistical, while the shaded bands indicate the systematic
    uncertainties of the results.  The contribution labelled
    "theor.~syst."  in the left panel gives the uncertainty due to
    with the use of two different unpolarised parton densities
    [\protect\refcite{lit:cteq5,lit:grv98}].
    }
  \vspace*{-1.5ex}
\end{figure}

The preliminary data presented here are based on 1.8 (6.5) million DIS
events taken on a hydrogen (deuterium) target in the kinematic range
$Q^2 > 1$ GeV$^2$ and $W^2 > 10$ GeV$^2$.  For the hydrogen data set, final
state pions could be identified using the information from a threshold
\v{C}erenkov counter, while for the deuterium data set the RICH detector
provided kaon identification as well.  The preliminary results for the
inclusive and semi--inclusive asymmetries on the deuterium target are shown
in Fig.~\ref{fig:A1d}.

Together with a set of inclusive and semi--inclusive asymmetries on a
polarised proton target, the polarisation of five quark flavours was
extracted according to Eq.~(\ref{eq:puri}).  For the sea quark
flavours, at values of $x > 0.3$ the polarisation was set to zero.
The resulting small influence on the up and down quark polarisations
was included in the systematic uncertainties.  Fig.~\ref{fig:dq5par}
shows the obtained quark polarisations as well as the polarised quark
distributions.  The polarisation of the up quarks is positive
everywhere in the measured range and increases with $x$ up to $0.7$ at
$x = 0.47$.  The down quark polarisation ranges between $-0.1$ and
$-0.5$, almost independently of $x$.  For the light sea quarks the
polarisation is compatible with zero, while for the strange quarks
a slightly positive polarisation is favoured within the measured range in
contrast to results based on inclusive data alone
(e.g.~[\refcite{lit:grsv2000,lit:bbfit}]).  However, within their
total uncertainty also the polarisation of the strange quarks is zero.

\begin{figure}[t]
  \vspace*{-1mm}
  \centerline{
    \includegraphics[width=0.52\textwidth]{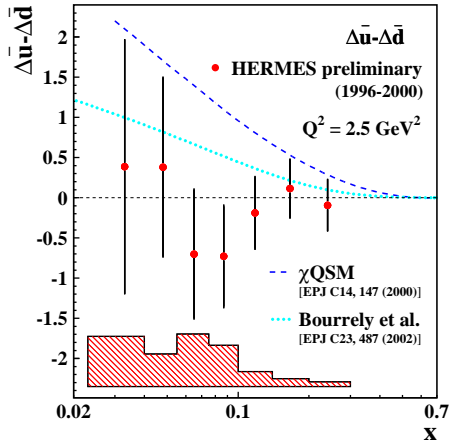}
    }
  \vspace*{-1mm}
  \caption{\label{fig:dubar-ddbar} HERMES preliminary result on the
    difference of the polarised light sea $\Delta \bar{u} - \Delta
    \bar{d}$ as a function of $x$ at a scale of $Q^2 = 2.5$ GeV$^2$.}
  \vspace*{-1.0ex}
\end{figure}

Fig.~\ref{fig:dubar-ddbar} shows a first measurement of the difference
of the polarised light sea flavours $\Delta\bar{u}(x) -
\Delta\bar{d}(x)$ together with predictions of a broken SU(2)$_f$
symmetry from Refs.~[\refcite{lit:goeke,lit:bourrely}].  The data are
consistent with an unbroken SU(2)$_f$ symmetry in the light polarised
sea.

\section{Conclusions}

HERMES has collected large statistics of inclusive and semi--inclusive
DIS data on pure targets of polarised hydrogen and deuterium.  From
the inclusive data precise results on the polarised structure function
$g_1^d$ were extracted.  Based on a LO QCD analysis of semi--inclusive
data taken on both target nuclei, polarised parton densities for $u$,
$\bar{u}$, $d$, $\bar{d}$, and $(s + \bar{s})$ flavours have been
extracted as a function of $x$ in the range $0.023 < x < 0.6$.  A
first measurement of the flavour symmetry in the polarised light sea
$\Delta\bar{u} - \Delta\bar{d}$ is compatible with the SU(2)$_f$
assumption.


\end{document}